\documentclass[aps,preprintnumbers,amsmath,amssymb,prl,superscriptaddress,longbibliography,twocolumn]{revtex4-2}
\usepackage{mwe}
\usepackage{dcolumn}
\usepackage{color}
\usepackage{xcolor}
\usepackage[caption=false]{subfig}
\usepackage{feynmp}
\usepackage{feynmp-auto}
\usepackage{float}
\usepackage{systeme,mathtools}
\usepackage{easyReview}
\usepackage{amsmath,amssymb,amsfonts,empheq,physics,float,mathrsfs,wrapfig,indentfirst,enumitem}
\usepackage{physics}
\usepackage{amsmath}

\def\comment#1{}

\newcommand{\nc}{\newcommand}

\nc{\scs}{\scriptstyle}
\nc{\setval}{\fmfset{wiggly_len}{3mm} \fmfset{arrow_len}{1.5mm}
	\fmfset{arrow_ang}{13} \fmfset{dash_len}{1.5mm}\fmfpen{0.125mm}
	\fmfset{dot_size}{2thick}}

\usepackage{bm,latexsym,mathrsfs,enumerate,color}
\usepackage[mathcal]{euscript}
\usepackage[breaklinks=true,unicode=true,urlcolor = blue,colorlinks = true,citecolor = blue,linkcolor = blue]{hyperref}
\usepackage{graphicx}
\usepackage{todonotes}
\usepackage{wrapfig}
\usepackage{easyReview}
\usepackage{mathbbol}
\usepackage{stmaryrd}

\renewcommand{\vec}[1]{\bm{#1}}

\def\slashchar#1{\setbox0=\hbox{$#1$}           
	\dimen0=\wd0                                 
	\setbox1=\hbox{/} \dimen1=\wd1               
	\ifdim\dimen0>\dimen1                        
	\rlap{\hbox to \dimen0{\hfil/\hfil}}      
	#1                                        
	\else                                        
	\rlap{\hbox to \dimen1{\hfil$#1$\hfil}}   
	/                                         
	\fi}                                         %

\DeclareMathAlphabet\mathbfcal{OMS}{cmsy}{b}{n}
\DeclareSymbolFontAlphabet{\amsmathbb}{AMSb}%

\begin{document}
\title{Hyperscaling of Fidelity and Operator Estimations in the Critical Manifold}
\author{Matheus H. Martins Costa}
\affiliation{Institute for Theoretical Solid State Physics, IFW Dresden, Helmholtzstr. 20, 01069 Dresden, Germany}

\author{Flavio S. Nogueira}
\affiliation{Institute for Theoretical Solid State Physics, IFW Dresden, Helmholtzstr. 20, 01069 Dresden, Germany}

\author{Jeroen van den Brink}
\affiliation{Institute for Theoretical Solid State Physics, IFW Dresden, Helmholtzstr. 20, 01069 Dresden, Germany}
\affiliation{Institute for Theoretical Physics and W\"urzburg-Dresden Cluster of Excellence ct.qmat, TU Dresden, 01069 Dresden, Germany}

\begin{abstract} 
    By formulating the renormalization group as a quantum channel acting on density matrices in Quantum Field Theories (QFTs), we show that ground-state expectation values of observables supported on slow momentum modes can be approximated by their averages on the fixed-point theories to which the QFTs flow. This is done by studying the fidelity between ground states of different QFTs and arriving at certain hyperscaling relations satisfied at criticality. Our results allow for a clear identification of cases in which one can replace a QFT by its scale-invariant limit in the calculation of expectation values, opening the way for a range of applications, including the improvement of numerical and analytical methods used to tackle the costly computer simulation of critical models.
\end{abstract}

\maketitle

\textit{Introduction} --- One central concept in the study of phase transitions and quantum field theories (QFTs) is that of a critical point: a point in the parameter space of the system, for field theories the space of couplings in the Lagrangian, at which the correlation length diverges, or equivalently, a first-order transition terminates \cite{stanley, goldenfeld, cardy_1996}. The modern understanding of this phenomenon is arrived at by studying the renormalization group (RG) and is encapsulated in Wilson's work \cite{wilson}. Focusing our attention on QFTs at zero temperature, the RG consists of describing the field theory as a flow in the space of couplings as we move from short to large scales. This dynamical flow has several fixed points, with their own directions of attraction and repulsion (called irrelevant and relevant, respectively) in parameter space, with the critical points, whose set define the ``critical manifold", associated to any saddle or node being those which flow towards it.

From this perspective, properties of field theories and phase transitions such as universality are elegantly explained through the fact that, for two theories flowing to the same fixed point, their correlation functions will quickly approach the same scale-invariant behavior in the infrared (IR) and thus become indistinguishable \cite{stanley, goldenfeld, cardy_1996, wilson, POLCHINSKI1984269}. However, this picture still leaves several questions unanswered: Is it possible to quantify the meaning of ``quickly approaching the same behavior"? And for which classes of observables can we guarantee that their expectation values in the critical QFT may be replaced by those at the fixed point?

These are not minor details, as, physically, a field theory at criticality can still be distinguished from a fixed point by measuring certain observables (such as those capable of resolving the unit cell in a solid), so there must be a quantitative criterion characterizing those which exclusively detect scale-invariant physics.

Furthermore, if we are able to identify such operators, this might prove of great use to the ongoing development of numerical simulations of QFT ground states. Such methods typically suffer from high computational costs for QFTs at criticality, due to long-range correlations \cite{mpsreview}. Then, were it possible to show that certain sets of observables display, effectively, the same expectation values in the actual critical theory and in the RG fixed point, this could help with the computational investigation of these systems through the use of powerful results applicable only to scale-invariant QFTs such as the conformal bootstrap \cite{conformalreview}, or the separability in momentum-space of their ground states \cite{paper2}.

Thus, the goal of this work is to answer the questions given above. We show that, for any margin of error $\epsilon$, there is a scale $\mu_\epsilon$ such that the difference between the expectation value of observables acting on scales smaller than $\mu_\epsilon$ calculated in the full ground state of the critical field theory and its IR fixed point is smaller than $\epsilon$. 

We arrive at this conclusion by applying the quantum informational concept of fidelity, see \footnote{Chapter 9 of the book \cite{nielsen} treats the topic in detail and many of the results described there are used in this work.}, and, in particular, its realization in QFTs, where it is better defined in terms of a logarithmic fidelity per unit volume \cite{fidelityvidal, fidelitygu, fidelityorderparameter}. In this way, we will also derive ``hyperscaling" relations satisfied by a local version of the fidelity between the fixed-point ground state and the density matrix $\rho_t$ generated by the Wilsonian quantum channel constructed in \cite{paper2} at the RG time $t$, as well as by $\mu_\epsilon$, a function of $\epsilon$. We conclude with a discussion of applications of our results, and further directions for this line of work.

Our analysis will assume that theories associated with the RG fixed points are Conformal Field Theories (CFTs). Since our focus is on unitary quantum systems, this is a rather weak assumption, see \cite{POLCHINSKIinv, rychkovscale, DiFrancesco:1997nk}, and it allows us to apply results such as the unitarity bounds \cite{conformalreview, Pal:2018idc} and operator product expansions \cite{opeconv}. Although the specific proofs given rely on these CFT properties, we suspect similar results would also apply to any scale-invariant, but not conformal, unitary QFTs.

\textit{RG quantum channel for critical theories} --- To begin, we first define precisely the objects of study. 

Following the formalism established in \cite{Balasubramanian:2011wt, paper1, paper2, paper3}, we can partition the Hilbert space of a given critical QFT in $d-1$ spatial dimensions as a tensor product of momentum modes $\mathcal{H} = \otimes_{\vec{k}}\mathcal{H}_{\vec{k}}$, where $\mathcal{H}_{\vec{k}}$ is defined as the subsystem of momentum $\vec{k}$ on which the Fourier-transformed field $\phi_{\vec{k}}$ acts. 

Now, field theories at criticality are characterized as perturbations of the RG fixed point by irrelevant operators \cite{wilson, goldenfeld, cardy_1996}. Furthermore, these perturbations must be regularized at some ultraviolet (UV) scale $\Lambda$, see \cite{weinberg_1995, weinberg_1996, Collins, conformalperturbation}, which for effective field theories is physically meaningful. The regularization procedure suitable for our purposes is given as follows: any scaling operator $\mathbb{O}_i$ may be formally written as a function of the field modes $\phi_{\vec{k}}$ (this comes, mathematically, from the irreducibility of the field operators \cite{pct, haag, cftwightman} and is made explicit in the path integral language \cite{jackiw, paper3}). Then, we define a regularized $\mathbb{O}_{\Lambda i}$ by setting to zero in $\mathbb{O}_i(\phi_{\vec{k}})$ all modes $\phi_{\vec{k}}$ with $|\vec{k}|>\Lambda$. 

A sharp momentum cutoff is implemented this way, which, as discussed in \cite{paper1, paper2, paper3} is key to the description of the renormalization group in terms of a momentum-space tensor product partition and becomes essential for a simple derivation of our results. This leads to (real space) non-localities in the effective action and in \cite{paper1} it is proven that they are responsible for the mixedness of the state $\rho_\mu$ generated when the fast modes are traced out. As will be shown, this makes the more common choice of smooth cutoffs \cite{Polchinski:1983gv, Bagnuls:2000ae, Goldman:2023gzr, goldman2024lindbladianexactrenormalizationdensity} inconvenient for our purposes. 

The quantities of interest in this work will depend on the correlation functions of these regularized scaling operators. In our scheme, this means that correlators of the form $\langle\mathbb{O}_{\Lambda,i}(\vec{x}_1,\tau_1)...\mathbb{O}_{\Lambda,i}(\vec{x}_n, \tau_n)\rangle_*$ are given by the Fourier transform of the CFT momentum-space correlation functions, calculated as in \cite{correlator1, correlator2}, with the restriction that the absolute value of spatial momenta is smaller than the cutoff.

Our choice of regularization may be somewhat abstract, however it is advantageous in the path integral formalism. By perturbing the fixed-point action $S_*$ to $S= S_* +\lambda_i \int_{x}\mathbb{O}_{\Lambda i}(x)$, we can integrate out all modes at scales above $\Lambda$ and arrive at the respective actions $S_{\Lambda}$, as well as, from our construction, $S_{\Lambda} +\lambda_i \int_{x}\mathbb{O}_{\Lambda i}(x)$. Moreover, following \cite{paper2}, the ground state of the fixed-point theory is separable when partitioned into momentum scales, that is $\ket{\Omega} = \otimes_\mu\ket{\Omega_\mu}$, where $\ket{\Omega_\mu}$ is an element of the space $\otimes_{|\vec{k}| = \mu}\mathcal{H}_{\vec{k}}$. Therefore, this guarantees that, when using the action $S_{\Lambda}$ to generate a density matrix via the path integral \cite{Balasubramanian:2011wt, nishioka, paper1}, we arrive at a pure state we denote $\rho_*$. Finally, we emphasize that, while certain choices of regularization for the theory and composite operators may be more or less convenient, their differences go to zero as the cutoff $\Lambda$ increases or universal behavior is approached, see \footnote{Chapters 6 and 7 of \cite{Collins} describe this in detail.}. 

Now the proper setup to describe the RG flow of QFTs at criticality in terms of their quantum states is established: given an irrelevant perturbation of an RG fixed point by a single operator $\mathbb{O}_i(\phi_{\vec{k}})$ with scaling dimension $\Delta_i>d$ (the generalization to multiple operators is simple), the critical QFT determines a state at scale $\Lambda$ by the action $S_{\Lambda} +\lambda_i \int_{x}\mathbb{O}_{\Lambda i}(x)$. 

The finite cutoff $\Lambda$ allows us to apply the Wilsonian quantum channel derived in \cite{paper2} and so, defining the RG time $t\coloneqq \log\left(\frac{\mu}{\Lambda}\right)$, where $\mu$ is the momentum scale separating ``fast" and ``slow" modes at a given step (with the former being the traced-out degrees of freedom), we have a family of states $\rho_t$, with $\lim_{t\to\infty}\rho_t = \rho_*$ as the critical theory flows from the UV to the IR. Meanwhile, from the perspective of the action functionals, we have the usual power law decay of the dimensionless coupling $g_i\coloneqq \lambda_i\Lambda^{\Delta_i-d}$ and other generated coefficients, which leads to a restoration of $S_{\Lambda}$ as $\mu\to0$ \cite{wilson, cardy_1996, goldenfeld}.

\textit{Fidelity and its infinite-volume limit} --- As previously stated, our goal is to derive estimations of the expectation values of certain observables in critical QFTs through the physical intuition that such field theories are ``close" or ``flow rapidly" to RG fixed points. In quantum information science, there is a rich literature dedicated to quantifying this notion, see \cite{nielsen, Bengtsson_Zyczkowski_2006} for an introduction. For us, the fidelity $F\left(\rho,\rho'\right) \coloneqq Tr{\sqrt{\sqrt{\rho}\rho'\sqrt{\rho}}}$ between two density matrices $\rho$ and $\rho'$, will be the relevant quantity. 

One can motivate the definition of the fidelity from many perspectives, but, for this work, two of its properties, discussed in \cite{nielsen}, will be key. First, if the density matrix $\rho'$ is a pure state $\ket{\Omega}\bra{\Omega}$, the expression for the fidelity reduces to
\begin{equation}
\label{fidex}
    F\left(\rho,\rho'\right) = F\left(\rho,  \ket{\Omega}\bra{\Omega}\right) = \sqrt{\bra{\Omega}\rho\ket{\Omega}}.
\end{equation}
That is, instead of a nonlinear formula requiring multiple matrix diagonalizations, we obtain an expectation value that can be handled with ease. This why the choice of a sharp RG cutoff is advantageous: the results of \cite{paper1,paper2} guarantee that the state $\rho_{\Lambda}$ defined previously is pure and we can use the simplified formula for the fidelity, whereas a smooth cutoff requires the introduction of real space non-localities in the quadratic term of the action by hand, making a systematic study $F\left(\rho,\rho'\right)$ with the new state $\rho_{\Lambda}'$ thus defined more difficult. On grounds of universality, we do not expect our main conclusions to be altered in the infrared, but prefer to use the sharp cutoff.

Secondly, we will make use of the fact that the fidelity provides an upper bound to the trace distance between $\rho$ and $\rho'$, defined as $D_{tr}\left(\rho,\rho'\right)= Tr\left|\rho-\rho'\right|$, via the inequality
\begin{equation}
\label{distineq}
    D_{tr}\left(\rho,\rho'\right)\leq 2\sqrt{1-F\left(\rho,\rho'\right)^2}.
\end{equation}
Moreover, the trace distance itself bounds the difference between the expectation values of an observable $\mathbb{E}$ in the state $\rho$ and the state $\rho'$. Taking $\mathbb{E}$ as a bounded operator (as all experimental measurements are), with largest eigenvalue $\lambda_{max}=1$ via a rescaling of the possible measurement results, we have \cite{nielsen, takesaki1}
\begin{equation}
\label{bound}
    \left|\langle\mathbb{E}\rangle_{\rho}-\langle\mathbb{E}\rangle_{\rho'}\right|\coloneqq \left|Tr\left[\mathbb{E}\left(\rho-\rho'\right)\right]\right|\leq D_{tr}\left(\rho,\rho'\right).
\end{equation}
As we are dealing with Quantum Field Theories, the observables we refer to should be thought of as bounded functions of the field operators once those are averaged over with a test (or fast-decreasing) function, as is standard in such cases \cite{haag, pct}. Moreover, observables acting only on modes below a scale $\mu$ will be bounded functions of the fields smeared with functions $f(\vec{x})$ whose Fourier transform $\hat{f}(\vec{p})$ has support restricted to a sphere of radius $\mu$ in momentum-space \cite{paper2}.

With this, the general idea is to show that the fidelity between the density matrix $\rho_t$ of a critical QFT along the RG flow and the (pure) state $\rho_*$ at the fixed point becomes approximately $e^{-\epsilon^2}$ at some time $t_\epsilon$, for any given $\epsilon\ll 1$. Whenever this happens, $D_{tr}\left(\rho,\rho'\right)$ is of order $\epsilon$, and so the difference between expectation values calculated with $\rho_t$ and $\rho_*$ will be within this accepted margin of error. Furthermore, as $\rho_t$ continues to approach the fixed point, for times larger than $t_\epsilon$ (scales below $\Lambda e^{-t_\epsilon}$), the error will be even smaller.

In order to implement this vision, however, we need to take into account some subtleties pertaining to the behavior of the fidelity in the thermodynamic/infinite volume limit of field theories. A general problem that appears when considering the fidelity $F$ between states of QFTs with different interactions is that, as the volume $V$ of the box used to regularize the theory is taken to infinity, $F$ tends to zero. In mathematical physics this is a result known as Haag's theorem \cite{pct, haag, Haag:1955ev, Earman:2006sdl}, while in condensed matter, Anderson recognized a similar mechanism under the name of ``infrared catastrophe" \cite{andersoninfrared}.

Physically, this phenomenon can be understood by considering the models at a finite volume. If, heuristically, we take the local states of two different QFTs, their fidelity will be some number $f_{loc}$ between $0$ and $1$. By translation invariance, we can roughly estimate the overlap of ground states as $F \approx (f_{loc})^{V}$, leading to an exponential decay of the fidelity as a function of $V$. This intuition is confirmed explicitly in certain models, see \cite{jackiw, duncan}, as well as the End Matter for a short derivation for critical QFTs. In fact, studies of the fidelity in relation to phase transitions have long recognized this and taken the physically meaningful quantity to be the logarithm of the fidelity per unit volume, at times called the fidelity susceptibility \cite{fidelityvidal, fidelitygu, fidelityorderparameter, zhoufidelity, fidelity2009, fidelity2016}. 

Such results lead to a conceptual tension that must be resolved: on one hand, since the fidelity between states of different QFTs is guaranteed to be zero, Eq. (\ref{bound}) seems to be unsuitable for providing any strong bounds. On the other, the physical notion that the \textit{local} states of theories with similar interaction couplings only have small differences suggests there should still be a way of bounding the error of expectation values of suitable observables.

We resolve this by focusing on localized operators, those acting on a region of total support volume $V_{s}$ and comparing their expectation values in the ground states of QFTs with Hamiltonians given by $H$ and $H+g\int_{\mathbb{R}^{d-1}}h_{int}$. For the perturbed Hamiltonian, the operator averages will not change by much if we replace the additional interaction term by $\int_{R}h_{int}$, where $R$ is a region containing the (effective) support of the operators and has total volume greater than, but of the order of, $V_{s}$. The reason for this is that, given the decay with distance of correlations between operators in QFT ground states \cite{pct, duncan, lieb, hastingscorr}, the removal of the region outside of $R$ in the interaction will produce errors that are suppressed as $V_s$ increases. Concretely, we show in the End Matter that the discrepancy between this estimation and the correct result is smaller than order $\mathcal{O}(V_{s}^{-1})$.

The insight, then, is that if the interaction $\int_{R}h_{int}$ reproduces the expectation values of interest with minimal error, we only need the fidelity between the ground states of the original Hamiltonian and such a truncated perturbation in order to use Eq. (\ref{bound}) as planned. As proven in the End Matter, this newly defined ``local fidelity" satisfies
\begin{equation}
\label{localfidelity}
    f_{loc}\left(\rho,\rho'\right) \approx \lim_{V\to\infty} \left(F_V\left(\rho,\rho'\right)\right)^{\frac{V_{s}}{V}}.
\end{equation}
Thus, we have a precise way of quantifying errors in the estimations of expectation values, as long as we remember that, as we change the support of the observable being analyzed, we must use a different $f_{loc}$, corresponding to the new $V_{s}$, in order to assess the margin of error. Note that with $f_{loc}$ we have provided an operational interpretation for the thermodynamic limit of the logarithmic fidelity density: it provides a sharp bound on the difference of averages of localized observables.

\textit{Hyperscaling of operator estimations} --- After the setup and conceptual discussions of the previous sections, we are now in a position to derive our main results and identify which observables of a field theory at criticality can be estimated using the IR fixed point ground state, up to a small error. 
The state $\rho_t$ generated by the RG flow at time $t$ reproduces exactly all correlation functions of observables $\mathbb{O}_\mu$ acting on the subsystem of modes below scale $\Lambda e^{-t}$, given by the Hilbert space $\otimes_{|\vec{k}|\leq\Lambda e^{-t}}\mathcal{H}_{\vec{k}}$, as long as we rescale $\mathbb{O}_\mu$ by a factor of $e^{-t}$, leading to a contraction of all its characteristic lengths, including its support \cite{Ma, paper2}. 

Using the description of the critical theory as the perturbation of a fixed-point by an irrelevant operator $\mathbb{O}_i$ and dimensionless coupling $g_i$, then, in the path-integral representation, $\rho_t$ is determined by an action functional of the form $S_\Lambda + g_i \Lambda^{d-\Delta_i} e^{-(\Delta_i-d)t}\int_x\mathbb{O}_{\Lambda i}(x)$ \cite{wilson, paper2}. 
Therefore, with the exponential suppression of interactions along the flow, the fidelity at a regularized finite volume $F\left(\rho_t,\rho_*\right)$ (and, by Eq. (\ref{localfidelity}) also the local fidelity) will rapidly converge to $1$, allowing us to derive bounds on estimation errors. As shown in the End Matter for a support volume $V_{s}$ we find
\begin{equation}
\label{logrg}
    \log f_{loc}\left(\rho_t,\rho_*\right) = - g_i^2e^{-2(\Delta_i-d)t}V_{s} e^{-(d-1)t}f(\mathbb{O}_{\Lambda i}),
\end{equation}
where $f(\mathbb{O}_{\Lambda i})$ is a function closely related to the fixed-point correlator $\langle\mathbb{O}_{\Lambda i}(x)\mathbb{O}_{\Lambda i}(y)\rangle$. A factor of $e^{-(d-1)t}$ appears multiplying $V_{s}$ due to the dilation transformation taking operators at scale $\mu$ back to the cutoff $\Lambda$, where we can compare the states $\rho_*$ and $ \rho_t$ and the expectation values they generate \cite{paper2}.

Now, in order to determine the local fidelity, one must calculate $f(\mathbb{O}_{\Lambda i})$ for each case of interest. However, by dimensional analysis, from the lack of units of $f_{loc}$ and the fact that in a critical QFT the UV cutoff $\Lambda$ is the only scale present, we easily conclude that $f(\mathbb{O}_{\Lambda i})  = C\Lambda^{d-1}$, with $C$ a numerical constant. The exact value of $C$ is not important, as it is $\mathcal{O}(1)$ and we are interested in the \textit{asymptotics} of the RG flow of the local fidelity, which thus depends only on universal information such as the exponent describing the decay of the irrelevant coupling.

Characterizing the volume $V_{s}$ of the support of different quasi-local observables in units of $\Lambda$, we define $v_s\coloneqq V_s\Lambda^{d-1}$. Thus, replacing $t= -\log\left(\frac{\mu}{\Lambda}\right)$ in Eq.(\ref{logrg}), we obtain the power-law dependence of the logarithmic local fidelity as a function of the scale $\mu$ below which momentum modes are being measured, as well as the typical size of the operator support in units of the inverse of the cutoff
\begin{equation}
    \log f_{loc}\left(\rho_t,\rho_*\right) \approx - \mu^{2\Delta_i-d-1}v_s.
\end{equation}
Note that the power-law dependence on $\mu$ has exponent equal to $2\Delta_i-d-1$ in a hyperscaling-like relation.

Finally, restoring the dimensionless coupling and cutoff in the equation above, if we demand that $\mu$ is such that $\log f_{loc}\left(\rho_t,\rho_*\right) \approx -\epsilon^2$ for $\epsilon\ll1$ (and so, the trace distance bounding differences between expectation values is approximately $\epsilon$ ), the scale $\mu_{\epsilon, v}$ from which critical expectation values of observables with support volume $v$ acting on slower momentum modes may be estimated using the fixed-point ground state with an error smaller than $\epsilon$ is given by
\begin{equation}
\label{scaling}
    \mu_{\epsilon, v} \approx \Lambda\left(\frac{\epsilon}{g_i\sqrt{v_s}}\right)^{\frac{2}{2\Delta_i-d-1}}.
\end{equation}

The hyperscaling exponent $\frac{2}{2\Delta_i-d-1}$ is guaranteed to be positive, as any irrelevant perturbation by definition satisfies $\Delta_i>d$. Thus, the estimation is consistent, in that it will not return a scale greater than the cutoff $\Lambda$, for any value of $\epsilon$. Note that if we perform this analysis for a nearly marginal perturbation, our results hold without further alterations, as the appearing logarithms can be resummed in the usual manner \cite{Collins} and the resulting $\Delta_i$, will still be greater than $d$. For \textit{exactly} marginal perturbations, the new QFT is a different fixed point of the RG and hyperscaling breaks down by the fact that none of the couplings flow anymore.

In order to see how our estimates work in practice, we can take the scaling dimensions of irrelevant operators in known fixed points and calculate a given $\mu_{\epsilon, v}$. For example, in \cite{exampleising} it was found that the irrelevant operator with lowest dimension for the 3d Ising CFT scales with exponent $\Delta\approx3.83$. If, for illustration purposes, our allowed margin of error is of $1\%$ and $g_{\epsilon'}$ is of order $\mathcal{O}(1)$, Eq. (\ref{scaling}) gives $\mu_{1\%}\approx4\times10^{-3}\Lambda$. As the usual crystal lattice has a spacing of $\approx0.1nm$, it means that measurements with resolution of $\approx 25 nm$ are already basically incapable of distinguishing a physical sample at criticality from the perfect scale-symmetric fixed point theory.

\textit{Conclusions and Outlook} --- We have shown that, although a QFT at criticality has a ground state different from the fixed point it flows to, the expectation values of a large class of operators can be calculated up to a very good accuracy by using the RG fixed point theory instead. Our work can be thought of as connected to the view \cite{lashkari1,lashkari2} of the renormalization group as an approximate error correction code \cite{lashkari1,lashkari2}, with the hyperscaling exponent providing the rate at which errors are corrected in the infrared.
Furthermore, we see our results as a tool to assist in the difficult problem of numerical simulations of critical QFTs. For instance, our analysis can contribute to ongoing investigations on the nature of ``deconfined quantum criticality" \cite{Senthil_2004, Wang:2017txt, senthil2023deconfinedquantumcriticalpoints}:

In \cite{so5multicriticalitytwodimensionalquantum}, signs are found that this transition is controlled by a multicritical fixed point with $SO(5)$ symmetry,  while at the same time there other proposals such as the ``walking scenario" from \cite{DeCesare:2025ukl, decesare2026disturbingnewsd2epsilonexpansion}. The hyperscaling relations can be used together with further computational calculations in order to calculate the expectation value of low-momentum observables in a lattice model flowing towards this tentative multicritical point \cite{so5multicriticalitytwodimensionalquantum} and comparing with conformal bootstrap calculations \cite{conformalreview, Chester:2023njo} of the same operator given in terms of the scaling fields. If the averages agree up to the error dictated by the hyperscaling of fidelity, it lends more credence to this explanation of deconfined criticality. On the other hand, if there are significant discrepancies, the expected fixed point is not present in the vicinity of the transition and other proposals must be given more weight. This approach is less computationally costly than the asymptotic scaling analysis of correlation functions required to obtain critical exponents, as the observables can be chosen to only act on a subset of the lattice which does not increase with the number of simulated sites.

Finally, while our focus has been on gapless critical theories, most, if not all, translation-invariant unitary QFTs flow to fixed points of the RG in the infrared, even those with an energy gap, whose IR limit is not necessarily trivial \cite{tqftreview, gukov2013topologicalquantumfieldtheory, Vojta:2023xzx, you2021fractoniccriticalpointproximate, Gorantla:2021bda, Lake:2021pdn}. Regardless, these RG fixed points still contain relevant and irrelevant directions in coupling space and, thus, ``critical manifolds".  It follows that the hyperscaling of fidelity, together with the results from \cite{paper2}, may be similarly useful to numerical studies of gapped QFTs flowing to such points (in fact, due to their gap and exponential decay of correlations, errors in the estimation of the local fidelity will be even more suppressed), a very interesting direction for future works.

\begin{acknowledgments}
	
 We thank the Deutsche Forschungsgemeinschaft (DFG) for support through the W\"urzburg-Dresden Cluster of Excellence on Complexity and Topology in Quantum Matter – ct.qmat (EXC 2147, Project No. 39085490) and the Collaborative Research Center SFB 1143 (project-id 247310070). 
 
\end{acknowledgments}

\bibliography{citations.bib}

\appendix
\section{End Matter - Calculations of the local fidelity}
\renewcommand{\theequation}{S.\arabic{equation}}

Here we prove some results used in the main text: the exponential decay with the volume of the fidelity between two states of different QFTs and the fact that, for localized observables, the ``local fidelity" can be justifiably used to estimate differences in expectation values. 

For conciseness, our derivations are made directly in the context of comparing a critical QFT with its IR fixed point, but from the arguments it will be clear that the conclusions apply to a vast family of field theories, excepting \textit{relevant} perturbations of fixed points, a case left for further investigation in the future.

We begin by explaining how we will calculate the fidelity between the states $\rho_*$ and $\rho_t$ in a finite volume $V$. As previously discussed, since $\rho_*$ is a pure state, $F_V(\rho_*,\rho_t) = \sqrt{\langle\rho_t\rangle_*}$ and we will find $\bra{\Omega}\rho_t\ket{\Omega}$ using the method developed in \cite{paper1}. Note that this requires a shift from zero to finite temperature and the use of CFT correlators as functions of $\beta$, at least in the limit it tends to infinity.

We make use of the fact that the matrix elements of $\rho_*$ and $\rho_t$ are given as the zero temperature limit of their finite temperature path integrals (with actions $S_*^\beta, S_t^\beta$, and partition functions $Z_*^\beta, Z_t^\beta$, respectively). With this, shifting some Euclidean time variables and manipulating the path integral, the overlap is given by

\begin{equation}
\label{pifidelity}
\langle\rho_t\rangle_* = \lim_{\beta\to\infty}\frac{1}{Z_*^\beta Z_t^\beta}\int_{2\beta}\mathcal{D}\phi\left[\left. e^{-S^\beta_*}\right\vert_0^\beta+\left. e^{-S^\beta_t}\right\vert_\beta^{2\beta}\right].
\end{equation}
With the vertical lines indicating the range of $\tau$ at which the respective finite temperature factors are non-zero \cite{paper1}. In practice these are implemented with Heaviside theta functions.

The idea now is to write $S_t = S_*+\lambda_i\int_x\mathbb{O}_{\Lambda,i}(x)$ (the decay of the coupling with $t$ being omitted for convenience) and expand both the exponential of the action and the partition function into a series in $\lambda_i$, leading to an expression dependent on fixed-point expectation values of the form $\langle\mathbb{O}_{\Lambda,i}(\vec{x}_1,\tau_1)...\mathbb{O}_{\Lambda,i}(\vec{x}_n, \tau_n)\rangle_*$.

As shown in \cite{paper1}, this construction has a number of properties: temporal translation symmetry is explicitly broken due to the change in the integrand at $\tau=\beta$ (meaning that when integrating over positions $\vec{x}$ and $\tau$, only an overall \textit{spatial} volume factor $V$ is produced), cancellations between terms generated from the exponential of the action and partition function occur such that the $\beta\to\infty$ limit is finite, and the final result is given by $1$ plus negative corrections starting at order $\mathcal{O}(\lambda_i^2)$. 

At each order in the expansion, the terms may be reorganized such that $F_V(\rho_*,\rho_t) \approx e^{-V\lambda_i^2f(\mathbb{O}_{\Lambda,i},\lambda_i)}$, where higher powers of $V$ appear from the disconnected components of the correlators (in our context of expanding around a CFT, this is the contribution of the identity operator in the OPE \cite{opeconv}). This procedure is consistent at every step and its convergence is that of Conformal Perturbation Theory \cite{conformalperturbation, conformalperturbation2} applied to irrelevant perturbations. And though the precise radius of convergence for the fidelity is unknown in this context, our relations can still be applied at any finite order of perturbation, as long as the correlation functions of the observables of interest are calculated at same order.

Now, with the exponential decay of the fidelity with system volume, we move on to the justification of the use of the ``local fidelity" and the form it takes. Consider, as stated in the main paper, the replacement of the interaction $\int_{\vec{x}}\mathbb{O}_{\Lambda,i}(\vec{x})$ by $\int_R\mathbb{O}_{\Lambda,i}(\vec{x})$, where $R$ is a region containing the support of some operator $A$ we wish to calculate the expectation value of and has volume $V_s$. 

The expectation value of $A$ in the new ground state associated with the modified interaction is given by the path integral expression

\begin{equation}
    \langle A\rangle'=\frac{1}{Z_*}\int\mathcal{D}\phi A e^{-S_*}\left[1-\lambda_i\int_{R}\mathbb{O}_{\Lambda,i}(x)+...\right].
\end{equation}

Writing $A$ as a function of the scaling operators at the fixed point \cite{pct, haag, cftwightman}, its expectation value is a combination of the CFT correlation functions involving $\mathbb{O}_i(x)$. For our purposes of error estimation, we need only to consider the example of $A$ equal to $\mathbb{O}_i(x)$ smeared with some function $g(\vec{x})$ of compact support, as all other correlation functions with $\mathbb{O}_i(x)$ have a faster power-law decay. Thus, 
\begin{equation}
    \langle A\rangle'\approx-\lambda_i\int_{\vec{y}}g(\vec{y})\int_{R,\tau}\langle\mathbb{O}_{\Lambda,i}(\vec{y},0)\mathbb{O}_{\Lambda,i}(\vec{x},\tau)\rangle. 
\end{equation}
Note that we would find the same expression for the expectation value $\langle A\rangle_t$ in the correct state $\rho_t$ by simply replacing the integration region of $\vec{x}$ from $R$ to the full regularized volume of the system.

If we choose $R$ such that the distance between its boundary and the support of $A$ is of the order of the length scale defining $V_s$ (such that $V_s$ differs from the volume of the support by some numerical factor which does not change our asymptotics), as $\mathbb{O}_i$ is an irrelevant operator and thus has scaling dimension $\Delta_i>d$, the difference between $\langle A\rangle'$ and $\langle A\rangle_t$ is, after a shift in positions, normalization of $g(\vec{y})$ and approximation of the integral (whose resulting error is of the same order of the estimate we will obtain), given by

\begin{multline}
     \langle A\rangle'-\langle A\rangle_t\approx \lambda_i\int d\tau\int_{\mathbb{R}^{d-1}\backslash R}d^{d-1}x\langle\mathbb{O}_{\Lambda,i}(0)\mathbb{O}_{\Lambda,i}(x)\rangle \\ 
     \approx \lambda_i\int d\tau\int_{\mathbb{R}^{d-1}\backslash R}d^{d-1}x\frac{1}{(\tau^2+\vec{x}^2)^{\Delta_i}}.
\end{multline}
Using the definition of region $R$ (which avoids points with $x\to0$ by construction) and estimating the integral through power counting arguments, if the length scale of $V_s$ is $l$, that is $V_s\approx l^{d-1}$,  we obtain

\begin{equation}
     \langle A\rangle'-\langle A\rangle_t\approx \frac{1}{l^{2\Delta_i-d}} < \mathcal{O}(V_s^{-1}).
\end{equation}
The reason we prefer to use the weaker bound $\mathcal{O}(V_s^{-1})$ is because this removes any dependence on the dimension $\Delta_i$ (for instance, if $\Delta_i\approx d$, the volume margin of error remains just as reliable as in other cases) and refers only to inherent properties of the operator being estimated. Finally, our derivations are easily generalized for more general observables $A$, as the resulting power law on $l$ is always bound from above by  $\mathcal{O}(V_s^{-1})$.

Such result means that the ground state of a fixed-point theory perturbed by an irrelevant operator acting on a bounded region correctly reproduces, up to errors of order smaller than $\mathcal{O}(V_s^{-1})$, the critical theory expectation values of local observables supported in this region. Therefore, for these specific operators, the local fidelity $f_{loc}$ between the fixed point and truncated perturbation can be used to bound errors in estimations.

Therefore, performing a calculation as in Eq. (\ref{pifidelity}), where now instead of the critical theory action $S_t^\beta$, we use the modified action $S^\beta_R$, we once again find an expansion containing the n-point correlation functions $\langle\mathbb{O}_{\Lambda,i}(\vec{x}_1,\tau_1)...\mathbb{O}_{\Lambda,i}(\vec{x}_n, \tau_n)\rangle_*$. However, as the factors of $\mathbb{O}_{\Lambda,i}$ come from $S^\beta_R$, the integrations over positions will only cover the region $R$ and, since the expectation values are with respect to the translation-invariant ground state $\rho_*$, will produce a volume factor of $V_s$. 

In more detail, considering the scaling properties of the regularized operator $\mathbb{O}_{\Lambda,i}$, as its fixed-point expectation value is zero, once again relevant results for the fidelity and local fidelity start at second order. From unitarity, it is easily seen that two-point functions of hermitian scaling operators (those which can perturb a fixed-point action) in an Euclidean CFT satisfy

\begin{multline}
    \int_Rd^{d-1}x_1\int_Rd^{d-1}x_2\langle\mathbb{O}_{\Lambda,i}(\vec{x}_1,\tau_1)\mathbb{O}_{\Lambda,i}(\vec{x}_n, \tau_2)\rangle_* \\
    < V_s\int_{\vec{x}}\langle\mathbb{O}_{\Lambda,i}(0,\tau_1)\mathbb{O}_{\Lambda,i}(\vec{x}, \tau_2)\rangle_*.
\end{multline}
The generalizations for higher-order correlation functions require a more detailed analysis of the terms added in order to arrive at spatial integrals such as those on the right-hand side. However, the higher-order correlators necessarily involve higher powers of the coupling and so these terms decay more strongly as we move to the IR, meaning the expression above is the only one needed.

From this inequality, we conclude that when the expansion in the coupling constant is summed to obtain the local fidelity $f_{loc}$, we can bound it from below by a function with the same exponential decay (actually, the same function $f(\mathbb{O}_{\Lambda, i}, \lambda_i)$), with the one difference that the regularizing volume $V$ is replaced with the support $V_s$. In other words, as claimed in the main text,

\begin{equation}
    f_{loc}(\rho_*,\rho_t) \approx \lim_{V\to\infty} \left(F_V(\rho_*,\rho_t)\right)^{\frac{V_s}{V}}.
\end{equation}

Furthermore, by restoring the dependence of $\lambda_i$ on the cutoff $\Lambda$ and RG time $t$, we may write for the evolution of $f_{loc}$ along the renormalization group flow in terms of the dimensionless factors $v_s$, $g_i$, $C$ (the latter a number of order $1$ from the finite temperature calculation) as
\begin{equation}
    \log f_{loc}\left(\rho_*,\rho_t\right) \approx - v_{s}Cg_i^2e^{-(2\Delta_i-d-1)t}.
\end{equation}
Plus terms with stronger exponential suppression.

With this, we have proven the properties of the local fidelity used to derive our results and can be confident on both the existence and value of hyperscaling exponent of the local fidelity, as well as the characterization of the set of operators from the critical theory which can be well estimated.

Lastly, we point out that our results do not change qualitatively when taking into account the corrections of order $\mathcal{O}(v_s^{-1})$ previously discussed, as well as the fact that low-momentum observables (acting only on modes below the cutoff) cannot be strictly localized in some bounded region of space. It is well-known that for ``quasi-local" operators \cite{haag}, for any small number $\delta$ (for us, taken much lesser than $\epsilon$), there will be a localized operator such that differences in matrix elements, and thus vacuum expectation values, between the two will be smaller than $\delta$. One, then, needs only to take the support volume $V_s$ such that the corresponding $\delta$ is sufficiently small.

Thus, our final expression for $\log f_{loc}\left(\rho_*,\rho_t\right)$ must include the corrections of $\mathcal{O}(v_s^{-1})$ and $\mathcal{O}(\delta)$, which will nevertheless be smaller than our margin of error $\epsilon$ for appropriately chosen low-momentum scale $\mu$ (which suppresses differences through the decaying coupling) and volume $V_s$ of the observable support.
\end{document}